\documentclass[aps,prb,twocolumn,floatfix,amsmath,amssymb,
superscriptaddress]{revtex4-2}
\usepackage{amsfonts}
\usepackage{stmaryrd}
\usepackage{bbm}
\usepackage{mathrsfs}
\usepackage{tipa}
\usepackage{amssymb}
\usepackage{txfonts}
\usepackage{graphicx}
\usepackage{dcolumn}
\usepackage{epstopdf}
\usepackage[colorlinks,linkcolor=blue,urlcolor=blue,citecolor=blue]{hyperref}
\usepackage{multirow}
\usepackage{subfigure}
\UseRawInputEncoding

\begin{document}
\newcommand*{\cm}{cm$^{-1}$\,}
\newcommand*{\Tc}{T$_c$\,}
\newcommand*{\DR}{$\Delta R/R$\,}
\newcommand*{\CVS}{CsV$_3$Sb$_5$\,}
\newcommand*{\CTB}{CsTi$_3$Bi$_5$\,}

\title{Infrared spectroscopy study of kagome material CsTi$_3$Bi$_5$}

\author{Liye Cao}
\affiliation{Center for Advanced Quantum Studies and School of Physics and Astronomy, Beijing Normal University, Beijing 100875, China}
\affiliation{Key Laboratory of Multiscale Spin Physics (Ministry of Education), Beijing Normal University, Beijing 100875, China}

\author{Xiangqi Liu}
\affiliation{School of Physical Science and Technology, ShanghaiTech University, Shanghai 201210, China}

\author{Jiayi Cheng}
\affiliation{Center for Advanced Quantum Studies and School of Physics and Astronomy, Beijing Normal University, Beijing 100875, China}
\affiliation{Key Laboratory of Multiscale Spin Physics (Ministry of Education), Beijing Normal University, Beijing 100875, China}

\author{Bixia Gao}
\affiliation{Center for Advanced Quantum Studies and School of Physics and Astronomy, Beijing Normal University, Beijing 100875, China}
\affiliation{Key Laboratory of Multiscale Spin Physics (Ministry of Education), Beijing Normal University, Beijing 100875, China}

\author{Xiaoting~Zhang}
\affiliation{Center for Advanced Quantum Studies and School of Physics and Astronomy, Beijing Normal University, Beijing 100875, China}
\affiliation{Key Laboratory of Multiscale Spin Physics (Ministry of Education), Beijing Normal University, Beijing 100875, China}

\author{Yanfeng Guo}
\affiliation{School of Physical Science and Technology, ShanghaiTech University, Shanghai 201210, China}
\affiliation{ShanghaiTech Laboratory for Topological Physics, Shanghai 201210, China}

\author{Fengjie Ma}
\email{fengjie.ma@bnu.edu.cn}
\affiliation{Center for Advanced Quantum Studies and School of Physics and Astronomy, Beijing Normal University, Beijing 100875, China}
\affiliation{Key Laboratory of Multiscale Spin Physics (Ministry of Education), Beijing Normal University, Beijing 100875, China}

\author{Rongyan Chen}
\email{rychen@bnu.edu.cn}
\affiliation{Center for Advanced Quantum Studies and School of Physics and Astronomy, Beijing Normal University, Beijing 100875, China}
\affiliation{Key Laboratory of Multiscale Spin Physics (Ministry of Education), Beijing Normal University, Beijing 100875, China}

\begin{abstract}

The kagome material CsTi$_3$Bi$_5$, which is isostructural to the extensively studied charge-density-wave (CDW) compound CsV$_3$Sb$_5$, exhibits intriguing electronic features within its two-dimensional kagome lattices of titanium atoms. Here, we perform optical spectroscopic measurements together with the first-principles calculations on single-crystalline CsTi$_3$Bi$_5$ to investigate its electronic properties comprehensively. It is found that the overall optical spectra exhibit surprisingly similarity to those of CsV$_3$Sb$_5$, even in the CDW state. Nevertheless, the existence of CDW instability is ruled out in CsTi$_3$Bi$_5$. Via careful comparison to the optical responses of CsV$_3$Sb$_5$, we attribute this difference to a significant reduction in the itinerant carrier density of CsTi$_3$Bi$_5$, which is associated with the absence of van Hove singularity near the Fermi level at $M$ point. These results support the scenario that the CDW in CsV$_3$Sb$_5$ is driven by the nesting of van Hove singularity. Additionally, we unveil some exotic low-lying absorption features, which provide clear evidence of flat bands in CsTi$_3$Bi$_5$. Our findings contribute to a deeper understanding of exotic phenomena in CsTi$_3$Bi$_5$ and provide valuable insights into the role of van Hove singularity in CsV$_3$Sb$_5$.

\end{abstract}

\maketitle
\section{introduction}

Recently, kagome material $A$V$_3$Sb$_5$ ($A$ = K, Rb, Cs) has attracted extensive attention in condensed matter physics. There are various exotic physical phenomena observed in the $A$V$_3$Sb$_5$ system, such as unconventional superconductivity (SC)~\cite{PhysRevLett.125.247002,Charge-order-and-sc}, charge-density-wave (CDW)~\cite{PhysRevLett.125.247002,Charge-order-and-sc,PhysRevX.11.031050,PhysRevX.11.041030}, pair density wave~\cite{pair-density-wave}, electronic nematicity~\cite{electronic-nematicity-CVS}, nontrivial topological surface state~\cite{HU2022495,PhysRevLett.125.247002}, anomalous Hall effect with the absence of magnetic order~\cite{FENG20211384,PhysRevB.104.L041103,PhysRevLett.125.247002} and so on. Such rich quantum phenomena are rooted in the unique electronic structure of $A$V$_3$Sb$_5$, which contains the general features of kagome lattice materials, such as flat bands~\cite{HU2022495}, Dirac cones with symmetry-protected Dirac points~\cite{PhysRevLett.125.247002,HU2022495}, and van Hove singularities (vHSs)~\cite{PhysRevB.104.L041101,10.1063/5.0081081,PhysRevB.104.L161112}. This system has provided an intriguing research platform for studying the interplay between SC, CDW order, and nontrivial topological electronic states.

Despite intensive investigations focused on $A$V$_3$Sb$_5$, there are still a number of intractable problems yet to be solved, particularly regarding the underlying mechanism of CDW order in \CVS. In this compound, some studies reveal the formation of a 2 $\times$ 2 $\times$ 2 superlattice in accompany with the CDW transition ~\cite{PhysRevLett.127.046401,pair-density-wave,PhysRevX.11.031026}, while others report the presence of a 2 $\times$ 2 $\times$ 4 superlattice~\cite{PhysRevResearch.5.L012032,PhysRevB.104.075148,4aCDW}. Moreover, anomalous Hall effect linked to time-reversal symmetry-breaking~\cite{PhysRevResearch.4.033145,CFP1} and electronic nematicity characterized by rotation symmetry-breaking~\cite{PhysRevB.106.205109,CDW-nematicity1,CDW-nematicity2} are observed in the CDW state, hinting at a possible influence of CDW order on symmetry-breaking and nontrivial topology~\cite{chiral-charge-order}. Fermi surface nesting is believed to play a crucial role in the development of CDW, as the vHSs are located very close to the Fermi level ($E_F$)  and could be perfectly connected by the CDW wave vector $q$~\cite{PhysRevLett.127.046401,PhysRevB.104.L041101,vHSs_CVS,VHS2}. However, the expected divergence of the electronic susceptibility corresponding to this Fermi surface nesting is not observed in theoretical calculation results~\cite{PhysRevB.105.045135}. At the same time, the effect of electron-phonon coupling is suggested to significantly contribute to the lattice distortion and thereby CDW instability~\cite{PhysRevB.105.L140501,PhysRevB.105.245121,EPC1}. Consequently, whether the origin of CDW stems from electron instabilities facilitated by vHSs~\cite{PhysRevLett.128.036402,PhysRevLett.127.046401} or lattice structure instability~\cite{PhysRevLett.127.046401} is not clear yet.

Investigating materials featuring analogous lattice structures as \CVS is beneficial to revealing the mysteries in this compound. Specifically, \CTB is a newly discovered kagome material that crystallizes in a layered structure with a space group $P$6/$mmm$, which is isostructural to \CVS~\cite{doi:10.34133/research.0238,ChinPhysLett.40.037102}. As revealed by experiments and density function theory (DFT) calculations~\cite{band}, the electronic structure of \CTB is highly similar to that of \CVS, except that the vHSs of \CTB are situated at about 0.15 eV and 0.75 eV above $E_F$~\cite{band,PhysRevLett.131.026701}, instead of being close to $E_F$ as in \CVS~\cite{vHSs_CVS}. Meanwhile, neither magnetic susceptibility, resistivity, or heat capacity measurements exhibit any discernible anomalies associated with CDW phase transitions in \CTB~\cite{measurements,ChinPhysLett.40.037102}, in contrast to the results of \CVS~\cite{PhysRevLett.125.247002}. This observation is further confirmed by subsequent studies, such as scanning tunneling microscopy (STM) and low-temperature x-ray diffraction measurements~\cite{electronic-nematicity,PhysRevLett.131.026701}. It is proposed that the absence of CDW order is related to the vHSs being pushed far away above the $E_F$ in \CTB~\cite{band,ChinPhysLett.40.037102}.

Furthermore, a superconducting transition at $\sim$ 4 K is reported in \CTB~\cite{yang2022titaniumbased}, although there are some controversies concerning this transition~\cite{measurements,ChinPhysLett.40.037102}. Electronic nematicity associated with orbital ordering is unveiled by some experiments compared with DFT calculations~\cite{electronic-nematicity,yang2022superconductivity,PhysRevLett.131.026701}, which is analogous to \CVS~\cite{CDW-nematicity1}. Further investigations are required to understand the underlying physics of these exotic phenomena thoroughly. Optical spectroscopy is a powerful technique for probing the charge dynamics and detecting possible gap-opening phase transitions. Here, we perform infrared spectroscopy measurements on single-crystalline \CTB from 300 K to 10 K, which reveal a very similar optical response to that of \CVS. The most prominent difference is that the broader Drude component of \CTB is greatly suppressed compared to \CVS, which is intimately related to the development of CDW in the latter. Additionally, we also observe some peculiar low-lying absorption characteristics at around 3100 \cm, which have no counterpart in \CVS. Combined with our DFT based first-principles calculations, these characteristics are attributed to interband transitions associated with flat bands about 0.35 eV below the $E_F$.

\section{methods}

Single crystals of \CTB were grown by using the flux method. The precursor CsBi was prepared by reacting Cs (purity 99.75\%) and Bi granules (purity 99.99\%) at 300 $^{\circ}$C. Starting materials of CsBi, Bi and Ti powder (purity 99.99\%) were loaded inside a Canfield crucible with a molar ratio of 3: 9: 1, which were then sealed in a quartz tube. The assembly was heated up to 1100 $^{\circ}$C in a furnace within 12 h and kept at this temperature for 10 h. It was subsequently slowly cooled down to 800 $^{\circ}$C at a temperature decreasing rate of 50~$^{\circ}$C/h and kept for 5 h, and then further slowly cooled down to 400 $^{\circ}$C at 2~$^{\circ}$C/h. Finally, the assembly was taken out from the furnace and decanted with a centrifuge to separate \CTB single crystal from the flux.

Infrared spectroscopy studies are performed on a Fourier transform infrared spectrometer (Bruker Vertex 80v) in the frequency range from 60 to 23 000 \cm. The single crystal of \CTB with a size of 5.0 $\times$ 6.0 $\times$ 0.8 mm$^3$ is used. The surface of this sample is newly cleaved in order to remove the oxide layer. An $in situ$ gold and aluminium overcoating technique is used to get the reflectivity $R(\omega)$. The real part of the optical conductivity $\sigma_1(\omega)$ is obtained by the Kramers-Kronig transformation of $R(\omega)$. The Hagen-Rubens relation is used for the low-frequency extrapolation of $R(\omega)$. We employ the x-ray atomic scattering functions in the high-energy side extrapolation~\cite{PhysRevB.91.035123}.

In our DFT calculations, the plane-wave basis based method and Quantum-ESPRESSO software package were used~\cite{QE2009, QE2017}. We adopted the generalized gradient approximation (GGA) of Perdew-Burke-Ernzerhof formula for the exchange-correlation potentials~\cite{perdew1996generalized}. Spin-orbit coupling is included in the simulations by using fully relativistic optimized norm-conserving Vanderbilt pseudopotentials from the Pseudo-Dojo library with a kinetic energy cutoff of 85 Ry~\cite{van2018pseudodojo,PhysRevB.88.085117}. A mesh of 12$\times$12$\times$9 k-points grid was used for sampling the Brillouin zone, and a denser uniform k-points grid of 20$\times$20$\times$20 was adopted for the non-self-consistent calculations of optical properties. During the simulations, all structural geometries were fully optimized to achieve the minimum energy.

\section{results and discussion}

\begin{figure}[t]
\centering
  \includegraphics[width=8.5cm]{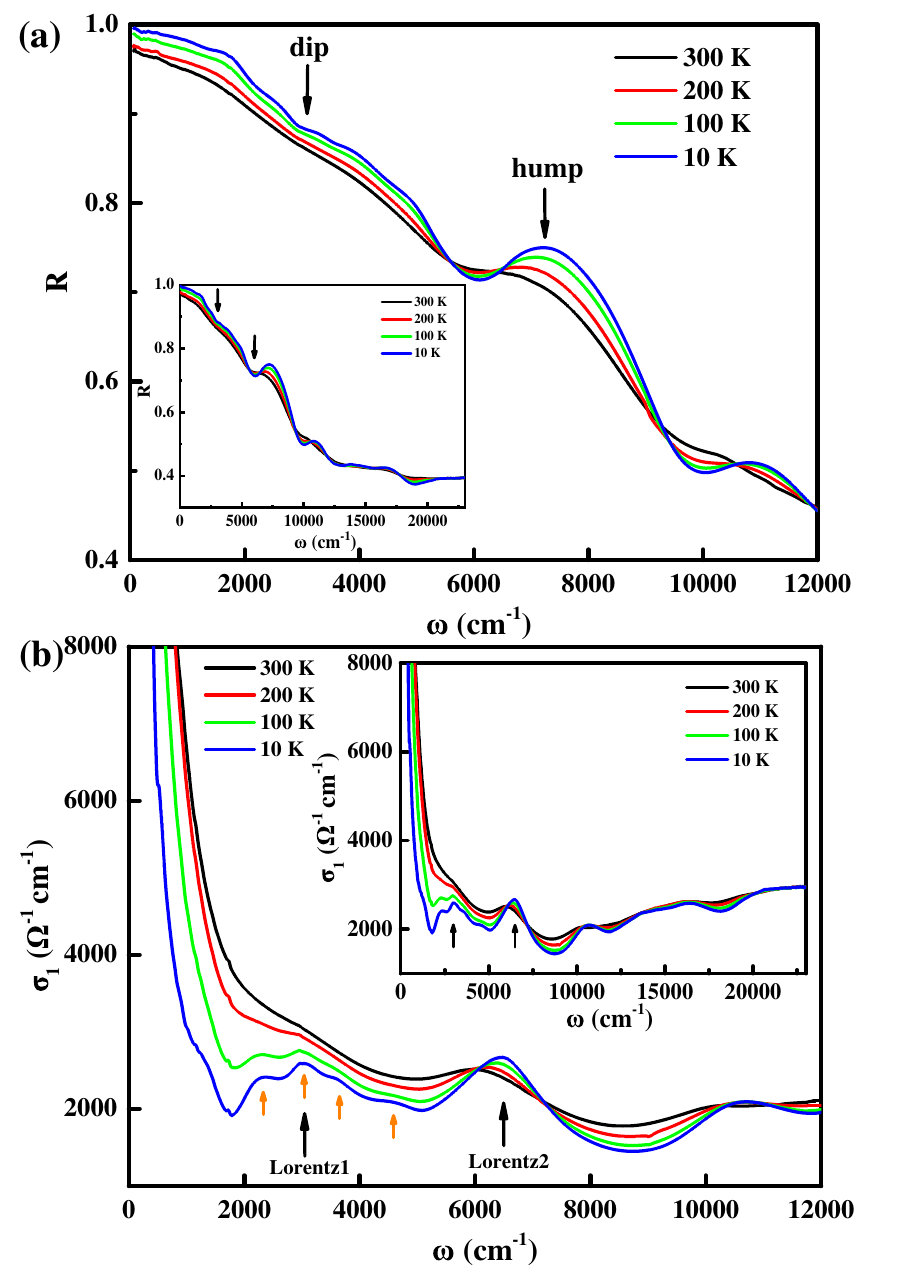}
  \caption{(a) The frequency-dependent reflectivity $R(\omega$) of \CTB. The inset shows $R(\omega$) in an expanded range up to 23 000 cm$^{-1}$. The black arrows indicate the position of the dip and hump. (b) The real part of optical conductivity $\sigma_1(\omega)$ of \CTB. The inset shows $\sigma_1(\omega)$ in an expanded range up to 23 000 cm$^{-1}$. The black arrows indicate the position of Lorentz1 and Lorentz2, and the orange arrows mark the position of several mini-peaks.}
  \label{Fig:R-sigma1}
\end{figure}

The reflectivity spectra $R(\omega)$ of \CTB at different temperatures are shown in the main panel of Fig.~\ref{Fig:R-sigma1}(a), while an expanded range up to 23 000 \cm is plotted in its inset. Across the entire measured temperature range, $R(\omega)$ increases as frequency decreases and approaches a unit at zero frequency, and the low-frequency reflectivity increases upon cooling, indicating a metallic nature of \CTB, which is consistent with previous reports~\cite{measurements,ChinPhysLett.40.037102}. 
Furthermore, the overall profile of the $R(\omega)$ spectra shows striking similarity to that of \CVS. In both compounds, the most pronounced feature below 10 000 \cm is a dip suppression together with a hump feature at higher energies, as marked by black arrows in Fig.~\ref{Fig:R-sigma1}(a). Moreover, these two characteristics get more pronounced with the temperature decreasing. It is worth to note that the development of suppression and enhancement of the hump in \CVS only appears below $T_{CDW}$, thus intimately related to the CDW phase transition~\cite{PhysRevB.104.L041101}. Therefore, it is tempting to explore whether the resembling dip and hump characteristics observed in \CTB are related to any kind of CDW condensate. 

On the other hand, there are some detailed differences between the $R(\omega)$ of \CTB and \CVS. Firstly, the suppression of \CVS is around 1500 \cm and the hump feature locates at around 6500 \cm~\cite{PhysRevB.104.L041101,PhysRevB.104.045130}.   In \CTB, the dip structure is much weaker and broader, which shifts to about 3000 \cm. Moreover, several even weaker mini-dips could be observed on top of the broad dip at 10 K and 100 K, which is absent in \CVS. 
Meanwhile the hump feature of \CTB locates at approximately 7200 \cm. Secondly, although the dip structure of \CTB seems to get more obvious with temperature decreasing, as expected for the CDW-related absorption, it is difficult to tell if there is an onset temperature where the absorption begins to show up. In addition, the hump feature is enhanced monotonically from 300 K to 10 K, which is unlikely to be related to any kind of phase transitions.   


To get a better perspective, we further resort to the real part of optical conductivity $\sigma_1(\omega)$ of \CTB, as shown in Fig.~\ref{Fig:R-sigma1}(b). It can be seen that the low-frequency $\sigma_1(\omega)$ consistently exhibits clear Drude components throughout the measured temperatures, i.e. low-energy peaks centered at zero energy, which reaffirms the metallic nature of \CTB~\cite{measurements,ChinPhysLett.40.037102}. The width at half maximum of the Drude peak represents the scattering rate of free carriers $\gamma$, which reduces monotonically with temperature decreasing, as can be seen in Fig.~\ref{Fig:R-sigma1}(b). In correlation with the dip and hump structures presented in $R(\omega)$, several Lorentz-type peaks show up in the optical conductivity of \CTB. The most interesting one is the one located at around 3100 \cm with a series of mini-peaks superposed on top, which corresponds to the dip structure donated in Fig.~\ref{Fig:R-sigma1}(a). For comparison, there is also a low-energy Lorentz peak in the optical conductivity of \CVS associated with the suppression of its reflectivity, which is identified to stem from the opening of a CDW gap. This similarity again puts forward the possibility of a CDW phase in \CTB.

\begin{figure}[tb]
\centering
  \includegraphics[width=8.5cm]{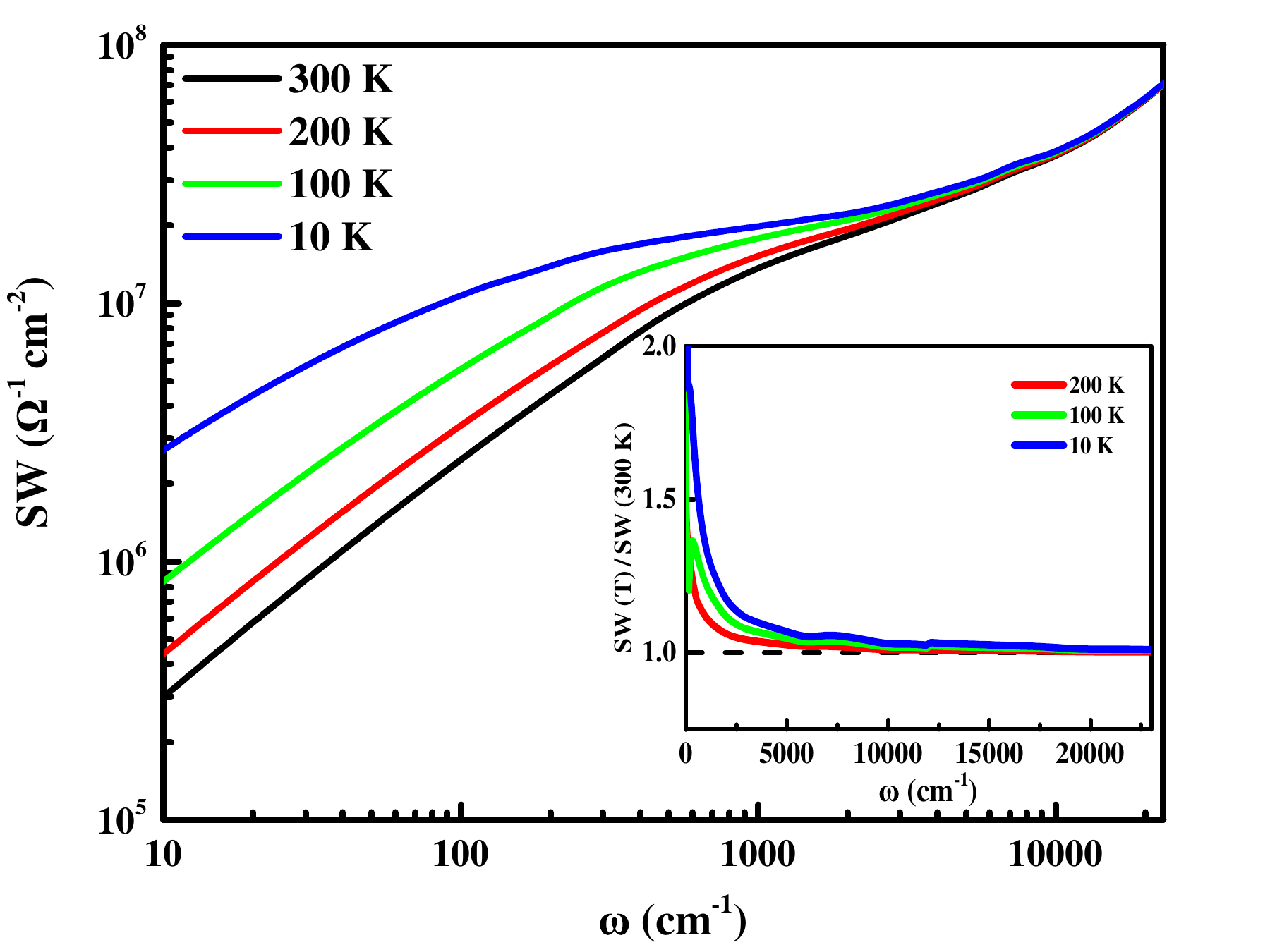}
  \caption{The spectral weight as a function of frequency. The insets: the renormalized spectral weight SW (T)/SW (300 K).}
  \label{Fig:SW}
\end{figure}

In infrared spectroscopy experiments, CDW orders manifest themselves in two complementary ways. The first one is the emergence of a Lorentz-type peak below $T_{CDW}$, whose central position shifts to higher energy and spectral weight (SW) piles up upon cooling. The second one is the transfer of SW from the itinerant Drude component to the emerging Lorentz peak, due to the loss of density of state accompanying the opening of the CDW gap. To prove the existence of a CDW transition, it is essential to provide evidence of these two manifestations. Here, it can be clearly seen from Fig.~\ref{Fig:R-sigma1}(b) that the 3100 \cm peak persists up all the way to room temperature, which suggests either no CDW transition or a CDW transition with $T_{CDW}$ higher than room temperature. As for the SW redistribution, we have plotted the SW as a function of frequency in Fig.~\ref{Fig:SW}. The SW generally increases with temperature increasing in a wide energy range and finally merges together roughly above 10 000 \cm. To further illustrate its relative change with temperature, we plot the SW(T) renormalized by SW(300 K) in the inset of Fig.~\ref{Fig:SW}. If there is a CDW-induced redistribution, the SW below $T_{CDW}$ is supposed to be smaller than that above $T_{CDW}$ at frequencies lower than the gap energy, which usually evidences as a dip structure in the renormalized SW~\cite{PhysRevB.83.155113,PhysRevB.106.245145,La3Ni2O7IR}. However, no such features are observed in the inset of Fig.~\ref{Fig:SW}. These results suggest that the CDW phase transition below room temperature is unlikely to occur in \CTB and the origin of the absorption characteristic around 3100 \cm is still unknown. 

\begin{figure}[htbp]
\centering
  \includegraphics[width=8.5cm]{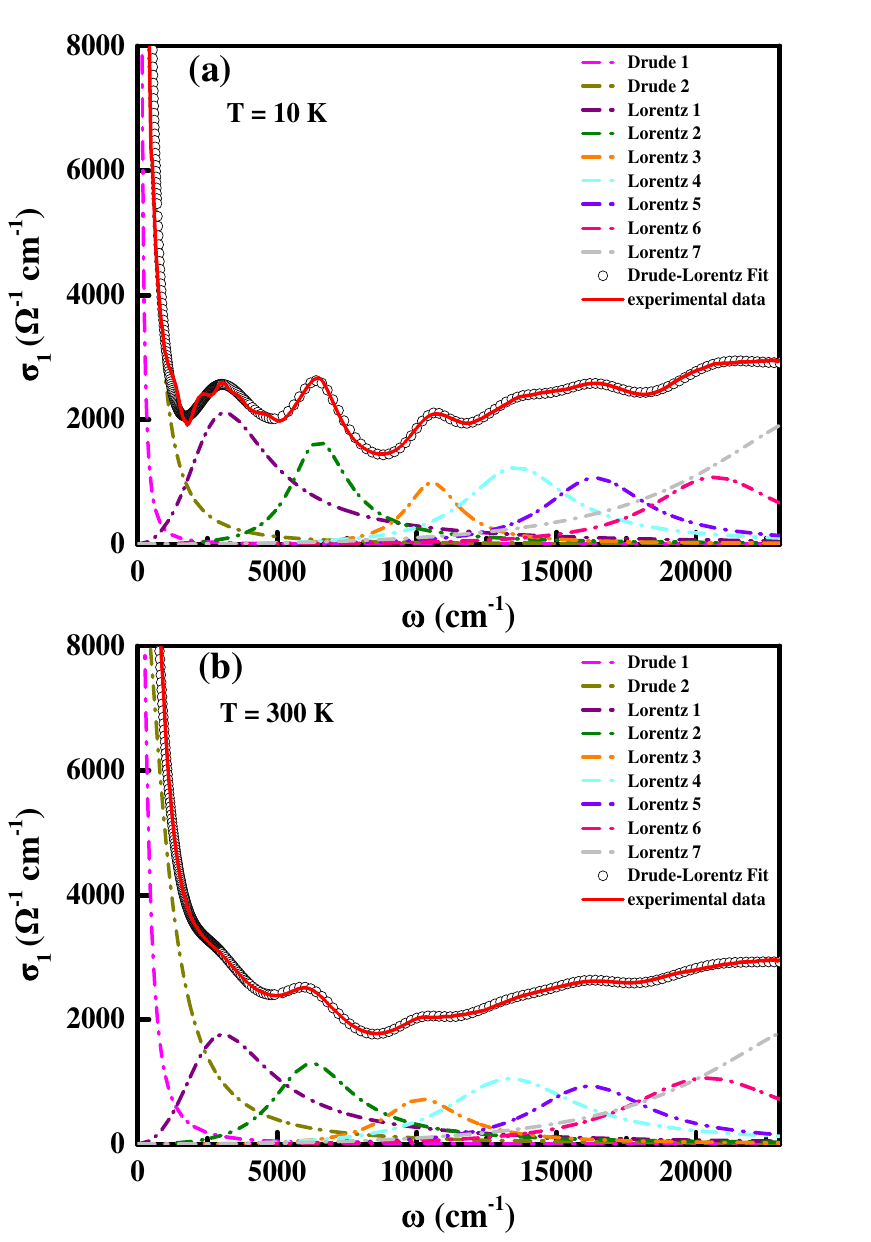}
  \caption{The fitting results of the real part of optical conductivity $\sigma_1(\omega)$ by Drude-Lorentz model at (a) 10 K and (b) 300 K of \CTB.}
  \label{Fig:3}
\end{figure}

\begin{table*}[htbp]
\normalsize
\caption{The fitting parameters at all measured temperatures in the unit of \cm. $\omega_{p,D_i}$ is the plasma frequency and $\gamma_{Di}$=1/$\tau_{Di}$ is the scattering rate of the Drude components. The resonance frequency $\omega_j$, width $\gamma_j$=1/$\tau_j$, and resonance strength $S_j$ of two Lorentz components are listed. The error bars of fitting parameters are included.
}

\setlength{\tabcolsep}{0.6mm}
\renewcommand\arraystretch{1.6}
\begin{tabular}{c|cccccccccccc}
  \hline
  \hline
  T & $\omega_{p,D_1}$ & $\gamma_{D1}$ & $\omega_{p,D_2}$ & $\gamma_{D2}$ & $\omega_1$ & $\gamma_1$ & $S_1$ & $\omega_2$ & $\gamma_2$ & $S_2$\\
  \hline
  10 K & 16219(283) & 52(1) & 21877(281) & 385(21) & 3128(207) & 3625(978) & 21412(2761) & 6487(227) & 2492(1285) & 15700(6588) \\
  100 K & 16142(226) & 124(2) & 23428(130) & 487(9) & 3116(43) & 3643(213) & 21245(621) & 6423(52) & 2678(304) & 15876(1570) \\
  200 K & 16083(309) & 251(5) & 24265(154) & 799(16) & 3085(26) & 3726(144) & 20377(453) & 6345(34) & 3142(203) & 16401(1034) \\
  300 K & 15748(180) & 285(3) & 24357(69) & 1038(18) & 3056(22) & 3878(131) & 20218(448) & 6232(32) & 3485(188) & 16452(993) \\
  \hline
  \hline
\end{tabular}
\label{TA:I}
\end{table*}

In order to further reveal the underlying mechanism of this exotic absorption, it is important to track down the quantitative variation of the charge dynamics. Therefore, we decompose the optical conductivity according to the Drude-Lorentz model,
\begin{equation}\label{Eq:1}
\sigma_1(\omega) = \sum_{i}\frac{\omega_{p,D_i}^2}{4\pi} \frac{\gamma_{Di}}{\omega^2+\gamma_{Di}^2} + \sum_{j}\frac{S_j^2}{4\pi} \frac{\gamma_j\omega^2}{(\omega_j^2-\omega^2)^2+\omega^2 \gamma_j^2}.
\end{equation}
It's noteworthy that this equation is structured in Gaussian units for the sake of convenience. The first term is the Drude components, which describe the charge dynamics of intraband carriers. The second term is the Lorentz components, which are used to describe excitations across the energy gaps or interband transitions. Here, $\omega_{p,D_i}=\sqrt{4\pi n e^2/m^{*}}$ and $\gamma_{Di} = 1/\tau_{Di} $ are the plasma frequency and the scattering rate of free carriers of the i${th}$ Drude component, where $n$ is the carrier density and $m^{*}$ is the effective mass. $\omega_j$, $\gamma_j = 1/\tau_j$, and $S_j$ are the resonance frequency, width, and resonance strength of the j${th}$ Lorentz component, respectively. To minimize fitting errors, we use the complex fitting method to fit the real parts of the optical conductivity using a nonlinear least-squares technique. The part of the fitting parameters with error bars at all measured temperatures are listed in Table~\ref{TA:I}. The representative examples of the fitting results of $\sigma_1(\omega)$ at 10 K and 300 K are plotted in Fig.~\ref{Fig:3}. Two Drude components with distinct scattering rates are demanded for the fitting, which indicates that the conduction electrons come from different energy bands or Fermi surfaces. Meanwhile, seven Lorentz terms are employed, although the latter five ones are almost temperature-independent and thus not listed. It is worth to remark that the mini-peaks around 3000 \cm are not taken into account in our fitting procedure. The peak positions of these mini-peaks are identified to be at 2362, 2972, 3630, and 4627 \cm at 10 K, as marked by orange arrows in Fig.~\ref{Fig:R-sigma1}(b).

\begin{figure}[tb]
\centering
  \includegraphics[width=9cm]{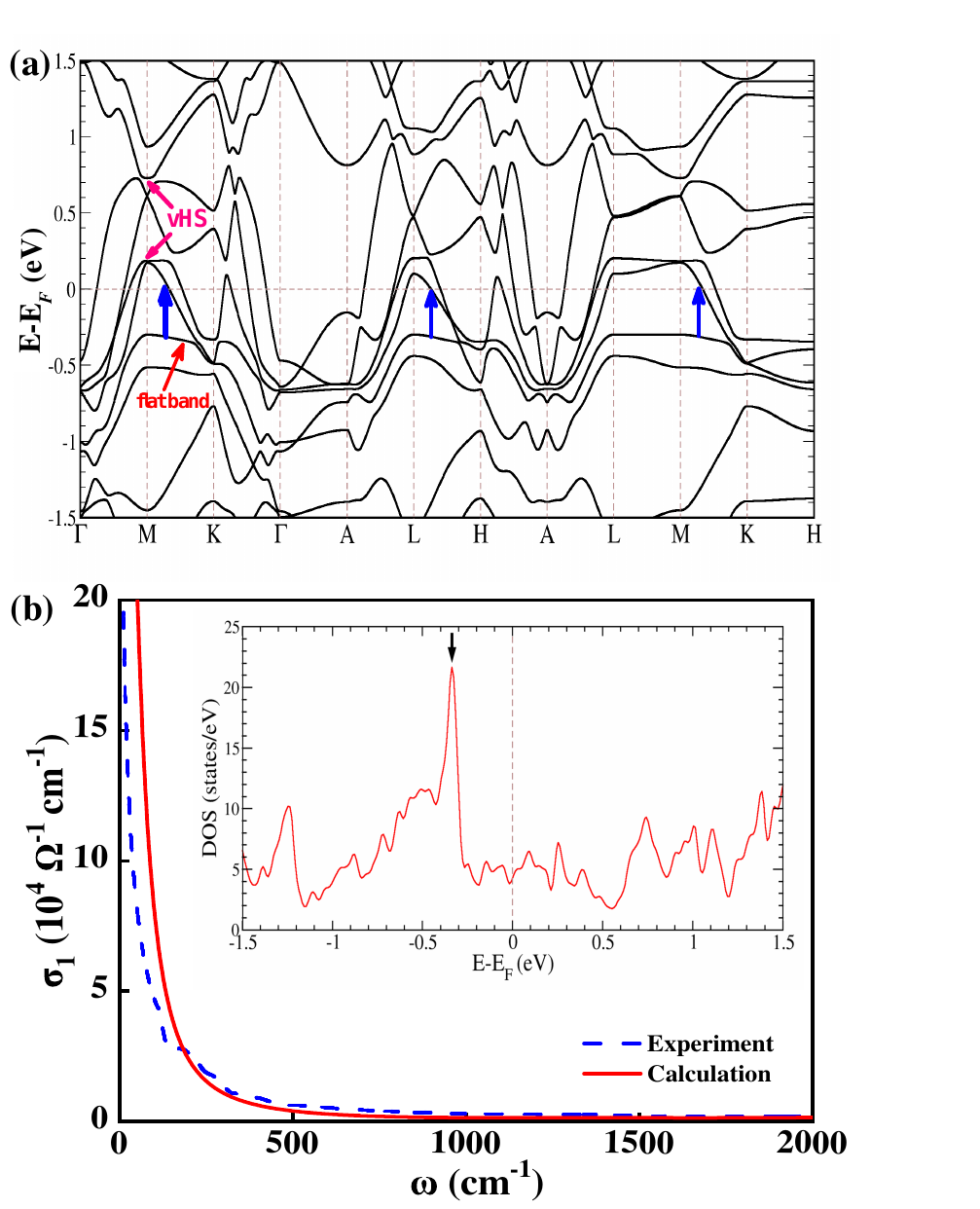}
  \caption{(a) The calculated band structure of \CTB. The vHSs and flat band are marked by pink and red arrows. The blue arrows indicate the interband transitions involving flat bands. (b) The measured $\sigma_1(\omega)$ (blue dashed curve) and the calculated $\sigma_1(\omega)$ (red solid curve) for \CTB. The inset shows the calculated density of states. The black arrow indicate the peak positioned at~$\sim$ 0.35 eV below the $E_F$.}
  \label{Fig:4}
\end{figure}

It is instructive to compare the charge dynamics of free carriers in \CTB with that in \CVS, where two Drude components with significantly different scattering rates are used as well \cite{PhysRevB.104.L041101}. We found that the scattering rates of \CTB generally decrease upon cooling, whereas the plasma frequencies are roughly temperature-independent. These characteristics are the same as \CVS when it is in the normal phase above $T_{CDW}$. Furthermore, the plasma frequency of the narrow Drude term ($\omega_{p,D_1}$) is roughly comparable between the two compounds, whereas $\omega_{p,D_2}$ of \CTB is much lower than that of \CVS~\cite{PhysRevB.104.L041101}, indicating a much smaller carrier density $n$ or much larger effective mass $m^{*}$. It is worth noting that for \CVS the narrow Drude term is ascribed to the light-electron-like and multiple Dirac bands at $\Gamma$ and $K$ points, whereas the broad one is attributed to the heavy hole bands near $M$ point. For comparison, we have calculated the electronic structure of \CTB by DFT. The band structure with spin-orbit coupling along the high-symmetry k-paths is presented in Fig.~\ref{Fig:4}(a). Similar to \CVS, there are also very light-electron bands crossing the $E_F$ around $\Gamma$ point, and Dirac type of conduction band near $K$ point. These bands usually contribute to Drude components with relatively small scattering rates~\cite{CVNbS-IR,RhSi-IR,TaAs-IR}, and thus highly likely to be responsible for the narrow Drude term (Drude1) of \CTB. In contrast, the band structure of \CTB at $M$ point is quite different from that of \CVS. The vHSs are pushed up to 0.15 eV and 0.75 eV above the $E_F$ instead of near the $E_F$, as indicated by the pink arrows in Fig.~\ref{Fig:4}(a). As a result, the density of states near the $E_F$ is expected to be much lower than \CVS. Assuming that Drude2 of \CTB is also related to the bands near $M$ point, this agrees perfectly with the fact that the $\omega_{p,D_2}$ of \CTB is much lower than \CVS. 

To discuss the possible contribution of effective mass $m^{*}$ to the suppression of $\omega_{p,D_2}$, it is useful to estimate the electron correlation strength, which could be reflected by the ratio of experimental kinetic energy $K_{exp}$ to theoretical kinetic energy $K_{band}$~\cite{PhysRevB.107.165123,La3Ni2O7IR,La4Ni3O10IR}. In simple metals with itinerant electrons of band mass, the agreement between the experimental and theoretical results would lead to $K_{exp}/K_{band}$ $\simeq$ 1. However, in correlated materials with quasiparticles of enhanced effective mass, the strong Coulomb interaction would drastically suppress the experimental kinetic energy, and thus generate a much smaller $K_{exp}/K_{band}$. The kinetic energy can be calculated by
\begin{equation}\label{Eq:2}
K = \frac{2\hbar^2c_0}{\pi e^2}\int_{0}^{\omega_c}\sigma_1(\omega)d\omega,
\end{equation}
where $c_0$ is the $c$-axis lattice parameter, and $\omega_c$ is the cutoff frequency covering the entire Drude components where $\sigma_1(\omega)$ reaches a minimum value. The theoretical $\sigma_1(\omega)$ calculated by DFT method using the random phase approximation is depicted by the red solid line in Fig.~\ref{Fig:4}(b). Here, $\omega_c$ = 1700 \cm is used to estimate both $K_{exp}$ and $K_{band}$, yielding a value of $K_{exp}/K_{band}$ = 0.74. This value is smaller than that of CsV$_3$Sb$_5$, which is about 0.81~\cite{PhysRevB.107.165123}. Therefore, a stronger correlation is expected in \CTB, which is consistent with the results of STM experiments~\cite{electronic-nematicity}. Nevertheless, this slight increase of correlation strength can hardly explain the substantially suppressed $\omega_{p,D_2}$ of \CTB compared to \CVS, which is therefore predominantly ascribed to the significant reduction of carrier density, or the shift of vHSs at $M$ point.

It seems that the optical responses of \CTB and \CVS are highly similar to each other, but this only stands in the normal states. Upon entering the CDW state, the broad Drude component of \CVS is significantly reduced, and the lost SW is transferred towards higher frequencies due to the formation of a CDW-related gap. In contrast, the SW of all the Drude and Lorentz terms of \CTB are stable against temperature variation, incompatible with a CDW phase transition. Particularly, the suspicious peak located at around 3100 \cm (Lorentz1) is actually robust up until 300 K, whose SW reduces very slightly with warming, as listed in Table~\ref{TA:I}. It only looks obscure at high temperatures in the $\sigma_1(\omega)$ spectra because it overlaps with the broadening Drude peak. Furthermore, the scattering rate of the broad Drude component in \CVS shows an abrupt reduction associated with the CDW transition temperature $T_{CDW}$~\cite{PhysRevB.104.L041101}, which is absent in \CTB. Therefore, we believe it is safe to conclude that there is no CDW-related physics in \CTB. Nonetheless, a detailed comparison between \CTB and \CVS would shed light on the origin of CDW in \CVS. We found the most prominent difference, from the spectroscopic view, lies in the SW of the broad Drude component, which are linked to the conduction bands near $M$ point. In other words, the itinerant carriers of \CTB are substantially reduced compared to \CVS, due to the absence of vHSs near the $E_F$. Therefore, our results strongly suggest that the CDW phase transition in \CVS originates from the vHSs near the $E_F$.

Finally, we would like to discuss the possible origination of the Lorentz1 peak centered at around 3100 \cm ($\sim$ 0.38 eV). Note that this peak does not have a counterpart in \CVS, and there are a series of mini-peaks superposed on it. In \CVS, the first Lorentz peak in the normal state is located at around 0.7 eV, which is argued to involve the electronic states near the vHSs at $M$ point. Since the vHSs of \CTB are significantly shifted upwards, they are unlikely to be responsible for the Lorentz1 peak of 0.38 eV. Instead, an obvious flat band is identified across the entire Brillouin zone in \CTB from our DFT results, which is in good agreement with ARPES measurements and previous calculations~\cite{ChinPhysLett.40.037102,band}. Especially, the flat band is located at around 0.35 eV below the $E_F$, which leads to a pronounced peak in the density of states at this energy, as indicated by a black arrow in the inset of Fig.~\ref{Fig:4}(b). Moreover, it closely matches with the peak position of Lorentz1 ($\sim$ 0.38 eV). Therefore, we propose that Lorentz1 corresponds to interband transitions from the flat bands to the unoccupied conduction band, as indicated by blue arrows in Fig.~\ref{Fig:4}(a). Additionally, slight dispersion of the flat bands could lead to additional tiny peaks on top of the main peak, as is also reported in some kagome materials, like Fe$_3$Sn$_2$~\cite{PhysRevLett.125.076403}. Hence, the mini-peaks on top of Lorentz1 provide clear evidence of the existence of flat bands in \CTB, which is a typical character of kogome materials, but not observed in \CVS yet.

\section{conclusion}

In conclusion, we have investigated the electronic properties of the kagome material \CTB through optical spectroscopic measurements and band structure calculations. Our findings reveal a metallic response in \CTB across the temperature range from 300 K to 10 K, consistent with its transport behaviors. Despite the striking similarity between the optical responses of \CTB and \CVS, there are no signatures of a CDW gap in \CTB. Moreover, the SW of the broad Drude component of \CTB is significantly suppressed compared to \CVS, which we attribute to the absence of vHSs near the $E_F$ in \CTB. This result supports the claim that the CDW formation in \CVS is linked to vHSs at $M$ point. Furthermore, the comparison between the experimental and theoretical optical conductivity of \CTB points to a little stronger electronic correlation than in \CVS. In addition, we have identified low-lying absorption features linked to interband transitions that involve flat bands in \CTB. Our work provides valuable insights into the underlying physics of exotic properties of \CTB and highlights the importance of vHSs in the development of CDW order in \CVS.

\begin{center}
\small{\textbf{ACKNOWLEDGMENTS}}
\end{center}
R. Y. Chen and F. J. Ma acknowledge the National Key Projects for Research and Development of China (Grant No. 2021YFA1400400), the National Natural Science Foundation of China (Grant Nos. 12074042 and 12074040), the Young Scientists Fund of the National Natural Science Foundation of China (Grant No. 11704033), and the Fundamental Research Funds for the Central Universities (Grant No. 2243300003). Y. F. Guo acknowledges the National Key R\&D Program of China (Grant No. 2023YFA1406100) and the Double First-Class Initiative Fund of ShanghaiTech University. This work was supported by the Synergetic Extreme Condition User Facility (SECUF). F. J. Ma was also supported by the BNU Tang Scholar.

\bibliography{ReferenceCsTi3Bi5.bib}

\end{document}